\documentclass[preprint,amsmath,amssymb,amssymb,superscriptaddress,aps,pra,nonbibnotes]{revtex4-1}
\usepackage{tabularx}
\usepackage{ragged2e}
\usepackage{graphicx,url}
\usepackage[utf8]{inputenc}  
\usepackage[english]{babel}
\usepackage{booktabs}
\usepackage{float}
\usepackage{url}
\usepackage{soul}
\usepackage{graphicx}
\usepackage{stmaryrd}
\usepackage{aeguill}
\usepackage{dcolumn}
\usepackage{bm}
\usepackage{braket}
\usepackage{times}
\usepackage{mathrsfs}
\usepackage{color}
\usepackage{xcolor}
\usepackage[pdftex, pdftitle={Article}, pdfauthor={Author}]{hyperref}
\usepackage{natbib}
\begin{document}
\title{Fast Design of Plasmonic Metasurfaces Enabled by Deep Learning}

\author{Abhishek Mall}
\affiliation{Department of Physics, Indian Institute of Technology -- Bombay, Mumbai - 400076, India}
\author{Abhijeet Patil}
\thanks{These two authors contributed equally}
\affiliation{Department of Electrical Engineering, Indian Institute of Technology -- Bombay, Mumbai - 400076, India}
\author{Dipesh Tamboli}
\thanks{These two authors contributed equally}
\affiliation{Department of Electrical Engineering, Indian Institute of Technology -- Bombay, Mumbai - 400076, India}
\author{Amit Sethi}
\email{asethi@iitb.ac.in}
\affiliation{Department of Electrical Engineering, Indian Institute of Technology -- Bombay, Mumbai - 400076, India}
\author{Anshuman Kumar}
\email{anshuman.kumar@iitb.ac.in}
\affiliation{Department of Physics, Indian Institute of Technology -- Bombay, Mumbai - 400076, India}

\begin{abstract}
Metasurfaces is an emerging field that enables the manipulation of light by an ultra-thin structure composed of sub-wavelength antennae and fulfills an important requirement for miniaturized optical elements. Finding a new design for a metasurface or optimizing an existing design for a desired functionality is a computationally expensive and time consuming process as it is based on an iterative process of trial and error. We propose a deep learning (DL) architecture dubbed bidirectional autoencoder for nanophotonic metasurface design via a template search methodology. In contrast with the earlier approaches based on DL, our methodology addresses optimization in the space of multiple metasurface topologies instead of just one, in order to tackle the one to many mapping problem of inverse design. We demonstrate the creation of a Geometry and Parameter Space Library (GPSL) of metasurface designs with their corresponding optical response using our DL model. This GPSL acts as a universal design and response space for the optimization. As an example application, we use our methodology to design a multi-band gap-plasmon based half-wave plate metasurface. Through this example, we demonstrate the power of our technique in addressing the non-uniqueness problem of common inverse design. Our network converges aptly to multiple metasurface topologies for the desired optical response with a low mean absolute error between desired optical response and the optical response of topologies searched. Our proposed technique would enable fast and accurate design and optimization of various kinds of metasurfaces with different functionalities. 


\end{abstract}

\maketitle
\section{Introduction}

Metasurfaces\cite{Yu2014,Kildishev2013,Chen2016} are ultra-thin arrays of nanostructures whose optical response can be engineered via the corresponding repeating units -- a meta-atom or a meta-molecule. These metasurfaces act as a compact and miniaturized platform for manipulating various properties of electromagnetic (EM) waves, such as polarization, amplitude, and phase. With advancements in nano-fabrication, these metasurfaces have led to applications in holography~\cite{li2019efficient}, beam splitting~\cite{niu2019photonic}, optical imaging~\cite{yang2014efficient}, sensing~\cite{lee2017metamaterials}, and miniaturized lensing~\cite{yin2017beam}. For designing these metasurfaces, traditionally multiple full wave electrodynamics simulations are carried out to optimize for the desired optical response or a functionality\cite{Achouri2018}. This process includes searching for either new non-intuitive nanostructures designs, or optimizing the geometrical parameters of a candidate nanostructure with simple design. Various EM Maxwell’s equations solver based brute force simulation methods are used to design and optimize nanostructures. Some examples include analytical modeling~\cite{molesky2018inverse}, evolutionary methods~\cite{wang2014broadband}, genetic algorithms~\cite{jafar2018adaptive} which involves an intelligent initial guess to reach a local optimum solution and numerical full-wave simulations using finite-element-method (FEM) and finite-difference time-domain (FDTD) with multiple parameter sweeps to search optimized values. However, these methodologies either suffer from significant expense of computational resources and time because of the iterative solving strategy or they simply fall short of capturing the functionalities with different-unique nanoantenna designs due to high non-linearity of nanophotonic design problem.

Alternatively, data driven structural optimization for directed functionality is a promising paradigm for solving complex computation problems. Deep learning (DL) is a subset of machine learning with gradient based optimization which is inspired by the human brain, where its logic, architecture, and functions are represented in the form of neural networks (NNs). In DL, a neural network learns the intricate correlation or mapping between inputs and outputs with minimum human intervention. It comprises of a stack of hidden layers composed of multiple neurons which is trained on a generalizable dataset. It captures the physical mechanisms statistically in terms of layered (or nested) but simple and well-defined non-linear functions. Lately, DL has been successfully employed in field of nanophotonics for device optimization~\cite{asano2018optimization,kiarashinejad2020knowledge,jiang2019global,qiu2019deep} and inverse design~\cite{long2019inverse,an2020freeform,an2019novel,ma2018deep,tahersima2018deep} for obtaining the desired optical response and directed functionality. While these works demonstrate the power of DL for nanophotonic device optimization with a one-time investment of a large EM simulation dataset, the inverse design process using DL still faces the issue of convergence to low error. This is either because of the one-to-many mapping where the same optical response can result from multiple geometrical designs and structural parameters, while on the other hand for certain desired optical responses, a design may not even exist. DL based generative models~\cite{liu2020hybrid,so2019designing,liu2018generative} or cascaded fully connected NNs~\cite{gao2019bidirectional,liu2018training,xu2019enhanced} have been employed for inverse design, but mostly they have complex NN architecture that lead to unstable results or require multiple models to be trained in parallel~\cite{ma2019probabilistic}.

To overcome the above-mentioned problems we propose a novel DL framework, which we call bidirectional autoencoder (biAE), for structural parameter optimization in \textit{multiple} metasurface topologies by successive parameter search of meta-atom or meta-molecule based metasurfaces to obtain a desired optical response. Our approach is fundamentally different compared to the previous works because it does not employ the inverse design process in the form of a pre-trained feed forward network. Our DL model, biAE, simultaneously learns forward and inverse mapping between the geometry of metasurface and corresponding spectrum. The simple architecture of biAE can be trained with a fewer number of EM simulation samples compared to other approaches ~\cite{peurifoy2018nanophotonic, so2019simultaneous}. Furthermore, we propose a novel loss function to train biAE which helps to capture sharp variations in the spectrum. Once our model is trained, it can generate accurate pairs of structural parameters in different geometric designs and corresponding optical response without the requirement of exhaustive computational resources.

To demonstrate the utility of the proposed modeling technique, DL architecture, and loss function, we focus on gap-plasmon based metasurfaces, where we implement polarization state conversion of reflected light by tailoring the geometry of the constituent nanoantennae~\cite{pors2015gap}. To generate the training data, appropriate combinations of these geometrical parameters were chosen by tuning several candidate designs for desired optical response. As examples, Figure~1(c) show the simulated conversion efficiency spectra for incident left circular polarization (LCP) to reflected right circular polarization (RCP) of gap-plasmon based half-wave plate metasurface (HMs) with a unit cell of a meta-atom and a meta-molecule. In this illustrative example, the framework searches for optimized parameters of different designs shown in Figure~1(b) of meta-atom (rectangle, double-arc) and meta-molecule (rectangle-circle pair, rectangle-square pair) in the GPSL generated using a proposed bidirectional autoencoder (biAE). We show the performance of our framework for the design of a single band, dual band, and a broadband gap-plasmon based half-wave plate metasurface with desired LCP to RCP conversion efficiency. Further, we show an example of how our template search methodology addresses non-uniqueness problem of metasurface design by searching $multiple$ different topologies for a desired optical response \textit{within an allowed error}, hence establishing a one-to-many mapping. This is important from the point of view of providing flexibility to the designer since the one optimal solution may not necessarily be the most useful, say in terms of fabrication constraints. In our example, we show that the search algorithm finds two geometrical designs from different design classes with optical response closely similar to the desired optical response with MAE measure of $\sim$ 2.5\%  (see Figure 7).

\section{Methodology}

We consider a simple metal-dielectric-metal periodic gap-plasmon based HMs~\cite{wu2017versatile} for demonstration of our method. The periodic nanostructure is capable of converting an incident wave of LCP to RCP and vice versa on reflection as shown in Figure~1 (a), where an unit cell with periodicity $P = 230$~nm which contains a meta-atom or a meta-molecule of Aluminium (Al) nanoantennae of height 50~nm. The Al-nanoantenna is mounted on a dielectric spacer SiO\textsubscript{2} layer over a 150~nm thick Al-mirror layer on the silicon substrate. Here we focus on different designs of Al-nanoantenna with variable parameters (see Figure~1) $L_x, L_y, D_s, D_B, w, v, a, d$, angle of rotation of nanoantenna ($\phi$) and SiO$_2$ spacer thickness $(t)$ which critically affects the optical response~\cite{pors2013efficient}. For this work, we fix the periodicity, thickness of Al-mirror layer, height of nanoantenna and silicon substrate thickness. We select the structural parameter range as mentioned in Table 1 to cover a broad range of conversion efficiencies.

A dataset comprising a total of 1500 samples is used for training and validation of the biAE. Each data sample is a pair of randomly selected structural parameters combination $s$ from different geometry design classes shown in Figure~1 (b) and the corresponding optical response $R$. The optical response is obtained as LCP to RCP conversion efficiency spectra from 400~nm to 800~nm wavelength range with 101 spectral points. In order to decide the adequate spectral resolution of the EM response in our training set we compare the different number of chosen spectral points (see Supplementary Information Figure 4) and find that the proposed biAE trained with 101 spectral points effectively captures original EM response and does not miss any feature. The data samples are derived from EM simulations performed using FEM in RF module of \textsc{comsol Multiphysics} with Livelink for \textsc{matlab}. The size of the dataset is decided so as to reduce the cost of the computation but at the same time reasonable enough to train the biAE consistently. A random split with 80$\%$ and 20$\%$ of the dataset is used for training and validation respectively for the biAE network. The dataset used contains optical responses with significant variability and multiple spectral features such as dips / peaks, oscillations and flat reflections for the defined set of  fixed and variable structural parameters for optimization (see Supplementary Information Figure 1). The variations in the training dataset help NN to learn generalizable features which improves the performance of the model significantly.
In stark contrast, learning from the dataset where optical responses have single resonance peaks or single kind of feature in transmission or reflection spectra becomes much convenient when trained on larger dataset fail to capture intrinsic features \cite{so2019simultaneous}. To circumvent this issue, bigger datasets with sufficient variations can be used for training deep learning frameworks. Generating these datasets using EM simulation is computationally expensive and time-consuming \cite{nadell2019deep, wiecha2019deep}. Another approach can be to use better deep learning algorithms which can learn from smaller datasets .  In the proposed method, we introduce simpler deep learning architecture in the form of biAE. We also introduce novel loss functions which help biAE to capture generalizable features from diverse dataset.

Figure~2 (a) shows a schematic illustration of the biAE model architecture, which was coded using the PyTorch library. The biAE consists of seven fully-connected layers having 101, 64, 32, 4, 32, 64, and 101 neurons, respectively. Each hidden layer used a rectified linear unit (ReLU) and Leaky ReLU non-linearity in encoder and decoder, respectively except for the final layer. The L$_{AE}$ objective function is the measure of performance of
the biAE during training and as the error
function during validation. Due to the exponential term, our objective function includes the role of a regularizer which is minimized by gradient descent using the Adam optimization algorithm in the training process. Batch normalization was used for fast convergence, while drop-out was used for regularization and to avoid over-fitting of the model. Adam optimizer was implemented to reduce the loss function. A learning rate of 0.0005 was used for training process. A dropout fraction of 0.5 was used for
each hidden layer. We monitored the difference between validation and training loss until both were stable and minimum. The L$_{AE}$ ( Equation 1), dropout and batch normalization ensured that we avoided over-fitting of the biAE. Our biAE takes the 101-dimensional optical response $(R)$ corresponding to structural parameters $(s)$ and encodes it to 4-dimensional structural parameters ($s^\prime$) as latent vectors using the encoder network. The output of the decoder $R^\prime$ is again a 101-dimensional vector, which emulates the input $R$ itself. The layer with the minimum number of neurons is the bottleneck of the biAE. It should be noted that we choose the number of bottleneck features to be four for all geometries because it fully captures geometry of nano-structures which significantly effects the optical response in this study. The number of bottleneck features can be easily changed according to complexity of geometry and structural parameters which can tailor the optical response with slight variations. Moreover, a dimensionality reduction approach could be applied on design space to decide the bottle neck size with features having strong correlations with EM optical response which also helps in further reduction in error for both forward and inverse design \cite{kiarashinejad2020deep}.

We optimize biAE to perform two tasks simultaneously. The model reconstructs the spectrum at the output of the decoder. Further, the encoder part of biAE tries to learn inverse mapping because of the loss constraints on bottleneck features. Hence the encoder and decoder network train simultaneously to reconstruct the optical response at decoder for an input optical response at encoder. The encoder network successfully encodes the optical response to structural parameters as latent vectors that allow the decoder to reconstruct the optical response and thus learning  correlations of inverse mapping (see Supplementary Information Figure 3).
Moreover, by minimizing the  $L_{AE}$ objective  function, the encoder network is forced to encode the features of EM response into a meaningful latent vector space (i.e structural parameters in our case) to facilitate the decoder with high reconstruction ability on EM response from structural parameters (demonstrated on the validation dataset in Supplementary Information Section G).
It is also important to note that our proposed biAE is different from the conventional conditional variational autoencoder network (CVAE), since we impose a MSE loss for bottleneck latent features to be close to geometrical design structural parameters. We do not impose any restriction on distribution of latent features as it is done in variational autoencoder \cite{NIPS2015_5775}. These choices of NN architecture and loss functions makes optimization for new dataset easier. As shown in Figure~5(a), multiple biAE could be trained with identical model architecture for more number $M$ of structural design classes. Here, we train four biAE of identical model architectures corresponding to our four different geometrical design classes, hence for our case $M = 4$ (see Figure 5). 

The training of the biAE with $N$ data points is completed by minimizing an objective function $L_{AE}$ using backpropagation. Our objective function is a combination of the exponential of absolute distance and the mean square error, as defined in Equation (1), which is strictly convex:

\begin{equation*}
 L_1= \left\{ \frac{1}{N}\sum_{i=1}^{N}(R' -R)^{2} \right\}+      \left\{\frac{1}{N}\sum_{i=1}^{N} \text{exp}(|R'-R|) -1 \right\}
\end{equation*}
\begin{equation}
L_2 = \frac{1}{N}\sum_{i=1}^{N}(s' -s)^{2},
\end{equation}
\begin{equation*}
L_{AE} = L_1 + L_2,
\end{equation*}
where $R$ is the optical response corresponding to structural parameter set ($s$), $R^\prime$ is the reconstructed optical response and $s^\prime$ is structural geometry latent vectors and $N$ is a mini-batch of training dataset. Adding exponential of absolute distance to MSE loss improves the performance of our model. This improvement is mainly observed when spectrum contains sharp variations because exponential function incurs more penalty when an error is large. After training of biAE, the optimized and trained decoder is used as an optical response generator for randomly selected structural parameters of different design classes with accurate correlations between $R$ and $R'$ indicated in Figure~2 (b). We also tried to train a standalone NN for forward and inverse design problem, in which case we found that the NN model tends to over-fit easily with high value of objective function during training phase  \cite{liu2018training, gao2019bidirectional}.  Here, the validation loss for decoder (L$_1$) was 0.00244  when trained using encoder as compared to 0.0177 when trained separately. Moreover the encoder losses (L$_2$) were 3.83 nm and 4.277 nm respectively for both cases. Therefore decoder training derived via encoder i.e a biAE, was more efficient than training separate designs for forward and inverse designs. Therefore when training biAE NN, it minimizes the error between the optical responses at input of encoder and output of decoder  and the problem at hand becomes the reconstruction of input rather than predicting the designs. This training felicitates the decoder on being able to train well using encoder. Moreover, to quantify the trade-off between the structural parameters encoded and EM response reconstructed, we train the network with objective function defined as L$_{AE}$ = L$_1$ + $\alpha$L$_2$, with $\alpha$ as a hyper-parameter ( see Supplementary Information Table 1) show that $\alpha=1$ gives the best performance on the validation dataset.

\section{Results and Discussion}

We validated a trained biAE with 100 data points from each geometry. These data points were not shown to biAE during the training phase. The data sample contains pairs of known randomly generated structural parameters and corresponding EM simulated spectrum. We input the EM simulated spectrum $R$ and obtain the regenerated spectra $R^\prime$ on the output layer. The results were compared with the EM simulated spectrum in the validation set, where we used the measure of average mean absolute error (MAE) per spectral point as given in Equation (3).

\begin{equation} 
L_{MAE} = \frac{1}{n}\sum_{i=1}^{n}|R'_i-R_i|
\end{equation}

The conversion efficiency regenerated by the biAE and corresponding EM simulated conversion efficiency for different meta-atom and meta-molecule geometrical designs for few randomly chosen validation samples is shown in Figure~3 (a)--(d). In the shown examples, the biAE regenerated conversion efficiency spectrum (red curves) matches the EM simulated (black curves) with good qualitative agreement giving an MAE of 1.43$\%$, 1.69$\%$, 2.66$\%$, and 1.90$\%$ for rectangle, double arc, rectangle-circle pair and rectangle-square pair respectively. The biAE neural network shows the capability to encode the high-dimensional optical response to low-dimensional structural parameters as latent vectors and retrieve the optical response back with high accuracy. For understanding the accuracy of our four biAE on the four geometrical design classes, in Figure~4 (a)--(d), we further plot average MAE on the validation dataset for each data sample of a geometrical design class with varying structural parameters. We observe that the average mean absolute error is $\leq 5\%$ for 97-100$\%$ of the dataset samples as shown in Figure~4. These results illustrate the ability of the biAE network to map the geometrical parameters to conversion efficiency with high precision. We conclude that the biAE successfully down-samples the high dimensional optical response to a low dimensional design parameters and reconstructs the optical response with up-sampling and does not lose the feature information while mapping.

With a small network architecture and a single NN optimized process, we demonstrate the effectiveness of our template search approach. We inverse design a gap-plasmon based meta-atom or meta-molecule for desired conversion efficiency optical response by accurately identifying best structural parameters combinations from our GPSL. The search for best parameters takes less than a millisecond, whereas full-wave simulation methodologies require at least few hours. Firstly, we generate a global library -- GPSL, of the entire conversion efficiency spectrum space given by all combinations of structural parameters with 10$^5$ entries from each geometrical design class using the decoder network as shown in Figure~5(a). The GPSL acts as a universal global library of geometrical designs and corresponding optical response comprising of sub-libraries (GPSL$_1$, GPSL$_2$, ....., GPSL$_M$) generated using the corresponding trained decoder networks. Here each sub-library indicates each metasurface class. The decoder network is highly efficient in generating optical response for randomly selected set of structural parameters from geometry design space. The generation of GPSL is feasible on Nvidia GeForce GTX 1050 GPU within a very short time. For a desired optical response $R^d$, the template search algorithm searches for an optical response with the least $L_1$ distance (depicted as $\Delta$) between desired optical response ($R^d$) and optical response ($R$) in GPSL and outputs the corresponding geometry design with structural parameter. The schematic of template search is shown in Figure 5 (b) . The template search algorithm is denoted as shown in Equation~ (4):
\begin{equation}
s_o = \arg\min_{s}| \Delta|, \  \text{and} \   \Delta =  R^d- R ,
\end{equation}
where $s_o$ is the optimized structural parameters searched, $s$ denotes the set of all structural parameters in the GPSL for all geometrical design classes of metasurfaces as discussed in section. The search algorithm outputs the design with least MAE from the list of $i$ designs with their corresponding $\Delta_i$. As examples of the results of the proposed method, we show the response of the searched structure from the library for a few desired conversion efficiency spectra in Figure~6. Here we choose desired spectrum as mixture of Gaussian to design a single band, dual band, and a broad band gap-plasmon based HMs of 40$\%$, 60$\%$-80$\%$, and 70$\%$ peak conversion efficiency. The Gaussian mixture is calculated as shown in Equation.~4:
\begin{equation}
g(\lambda)= \sum_m g_m \exp\{-\frac{(\lambda - \lambda_m)^2}{2\sigma_m^2} \},
\end{equation}
where $m$, $\lambda_m$ and $\lambda$ are the number of Gaussian, central wavelength of $m^{\text{th}}$ Gaussian and wavelength range respectively. For the optical response in Figure~6 (a)--(d), the search predicts rectangle, double arc, rectangle-circle pair and rectangle-square pair geometry design class with $s_o$ = [90, 199, 65, 90], [84, 104, 125, 135], [40, 170, 68, 150] and [70, 129, 51, 67] respectively for different desired optical response. We observe that our template search methodology shows no preference towards a specific design class and predicts different optimized geometries for various conversion efficiency ranges.

The training of the feed forward network for predicting an optical response from a design space is a fairly straightforward task. However, for the inverse design process any deep NN faces the issue of non-uniqueness --- multiple designs can have nearly identical optical responses. Hence, during training of the inverse design NN, existence of two or more outputs for a single input makes it hard to converge the NN training. To find the two or more optimized structural parameters $s_o^A$, $s_o^B$ etc. (if they exist) for a desired optical response $R^d$, we perform a template search in each sub-library corresponding to each geometrical design class. For every desired optical response as input, the template search algorithm searches a geometrical design with structural parameters which have least MAE ($\Delta$). Hence, while searching in sub-libraries, the algorithm outputs geometrical design with structural parameters from each design class. The algorithm results in the closest matching spectrum with least MAE from each sub-library. For instance, in Figure~7, we observe that there exist two optimized sets of geometry design for identical conversion efficiency optical response. The searched geometry designs are double arc as a meta-atom (red circle curve) and rectangle-circle pair as a meta-molecule (black circle curve) with $s_o^A=[92, 114, 65, 135]$, and $s_o^B = [65, 111, 72, 95]$ respectively, for a desired optical response (grey curve). The template search resulted in structural designs having similar optical response in comparison to desired input response with $\sim$ 2.5\% MAE. Essentially, the template search framework achieves the global minima while searching for design for a desired optical response in GPSL -- a global search; whereas it settles to local minima for each design class while attempting sub-library search-- a local search. This highlights the power of our technique in tackling the one-to-many problem of inverse design by performing a local and  a global search, where the proposed methodology by-passes non-convergence and searches for structural designs with reasonable allowed error. Subsequently, the user can manually chose the design most suitable for them from the point of fabrication constraints or otherwise. The predicted dimension of structural parameters are in range of fabrication using electron beam lithography \cite{Wu2016}. For a slight variation in a structural parameters during fabrication as compared to the predicted structural parameters by biAE will not effect the EM response of fabricated sample with a significant amount (see Supplementary Information Figure 5).

Furthermore, a simple application of template search algorithm on the dataset in comparison to a search on GPSL reveals the efficacy of biAE to model design and space correlations of nanophotonic metasurfaces, where the design space (the dataset) used to train biAE has data samples quite apart to do template search and obtain a closest matching optical response as a template search on GPSL response (see Supplementary Information Figure 2). This validates the fact that a NN model is indeed needed to learn the mapping between design and response space and could be used to design an universal library. A search methodology accompanied by efficiently trained NN model alleviates the need of training multiple models, tackles non-uniqueness problem during inverse design and can generate large design space for successive on-demand desired optical response search.

\section{Conclusion}

In this paper, we propose a DL framework based on a bidirectional autoencoder (biAE) paradigm for nanophotonic metasurface design via a template search methodology. Our methodology is distinguished from previous nanophotonic DL approaches in that it addresses optimization in the space of different metasurface topologies instead of just one, in order to tackle the one to many mapping problem of inverse design. As an example problem to demonstrate the power of our approach, we trained our framework to efficiently map the various topological designs of gap-plasmon based half-wave plate metasurfaces to the desired LCP to RCP conversion efficiency spectra. Our biAE learns to reconstruct high dimensional conversion efficiency spectra from low dimensional structural parameters through a bottleneck of low-dimensional latent vectors with average MAE $\leq$5$\%$ for 97-100$\%$ of validation data samples. The decoder network of trained biAE is subsequently utilized to create a Geometry $\&$ Parameters Space Library of conversion efficiency spectrum space given by all combinations of geometric parameters with $10^5$ data points for each of the different considered topologies of the metasurface. To verify the generality of the biAE based template search methodology, we performed a metasurface design search with multiple Gaussian as the desired spectra and obtained corresponding spectra from the GPSL that closely mimicked the desired spectrum. Our methodology addresses the non-uniqueness problem of nanophotonic inverse design by predicting multiple distinct geometry designs for a single desired spectrum and can easily be extended to the design of other kinds of metasurfaces and metamaterials \cite{So2020}.

\section*{Funding}
AK acknowledges funding support from the Department of Science and
Technology via grant numbers SB/S2/RJN-110/2017, DST/NM/NS-2018/49 and ECR/2018/001485.

\newpage
\begin{table*}[t]
\caption{Structural parameters values for different geometrical designs. All the distances are in nanometers and $\phi$ values are in degrees.}
\begin{tabular*}{\textwidth}{c @{\extracolsep{\fill}} cccccccc} \hline \centering
$L_x,L_y$ & $D_s$ & $D_B$ & $v$ & $w$ & $a$ & $d$ & $t$ & $\phi$ \\
\hline\centering
50-200 & 80-110 & 100-160 & 100-200 & 30-70 &
20-60 & 40-100 & 50-150 & 0$^0$,45$^0$,90$^0$,135$^0$ \\
\hline\centering
\end{tabular*}
\end{table*}

\begin{figure}[h!]
\includegraphics[width=5.75cm]{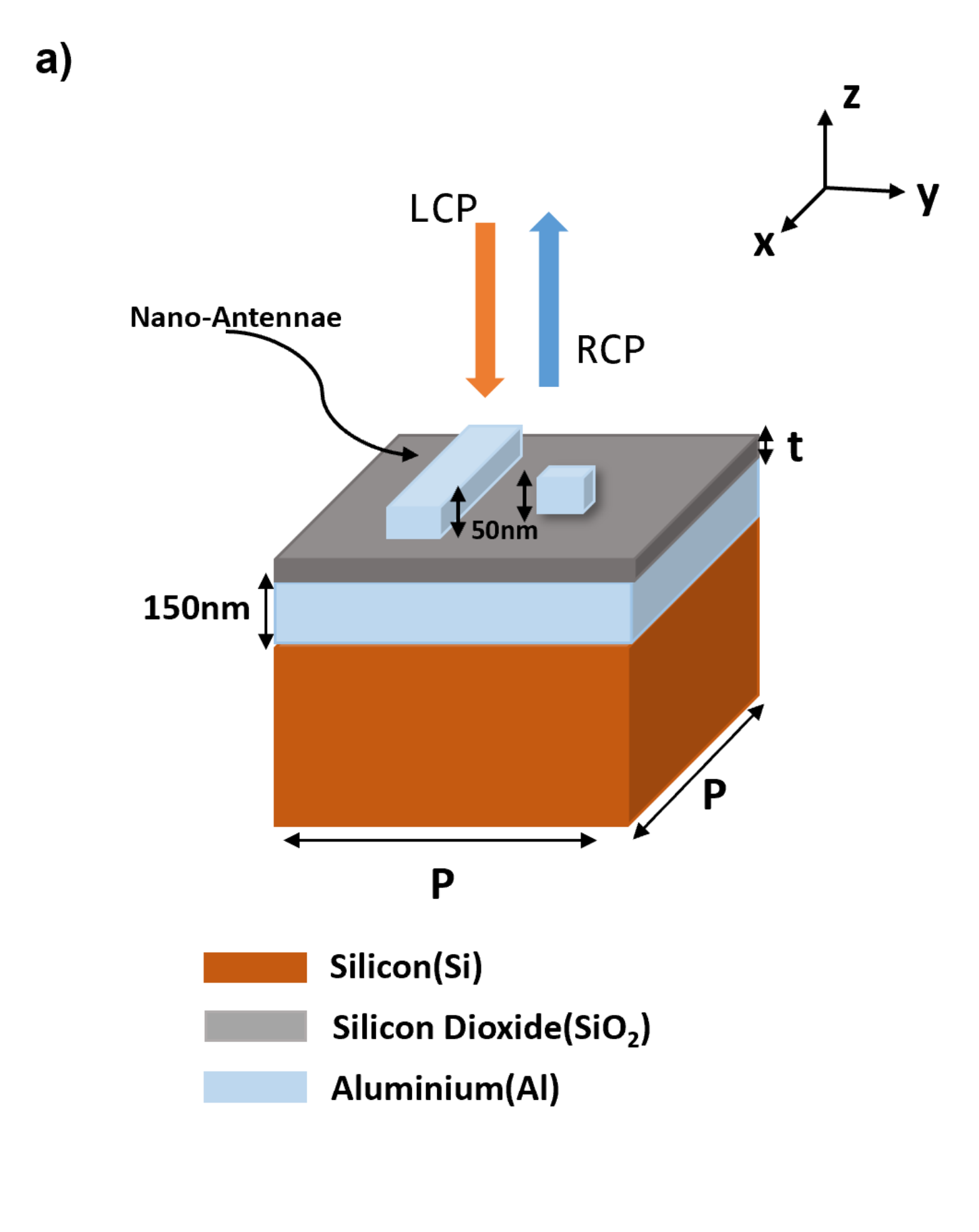}  
\includegraphics[width=7.1cm]{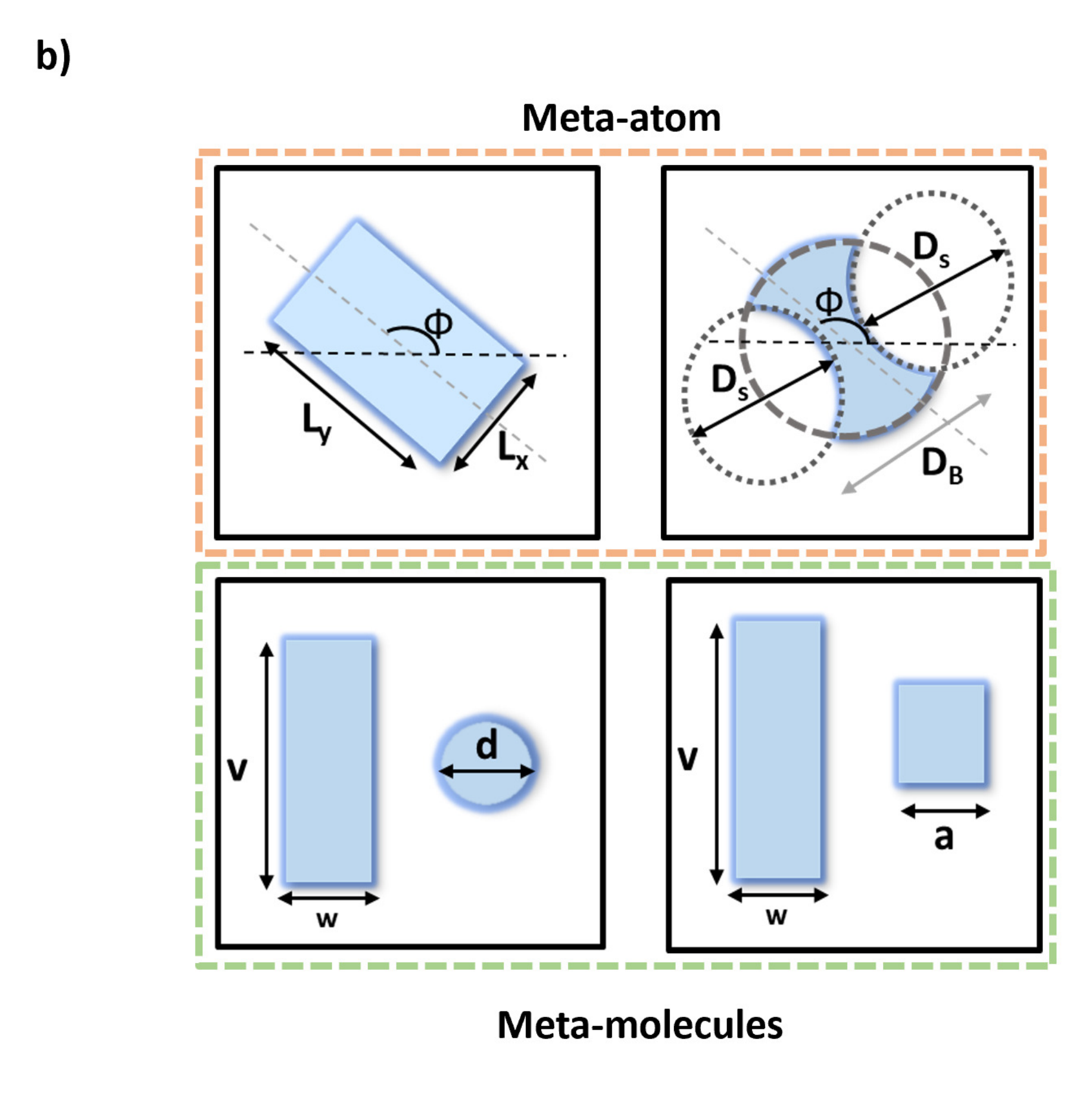}   
\includegraphics[width=6.7cm]{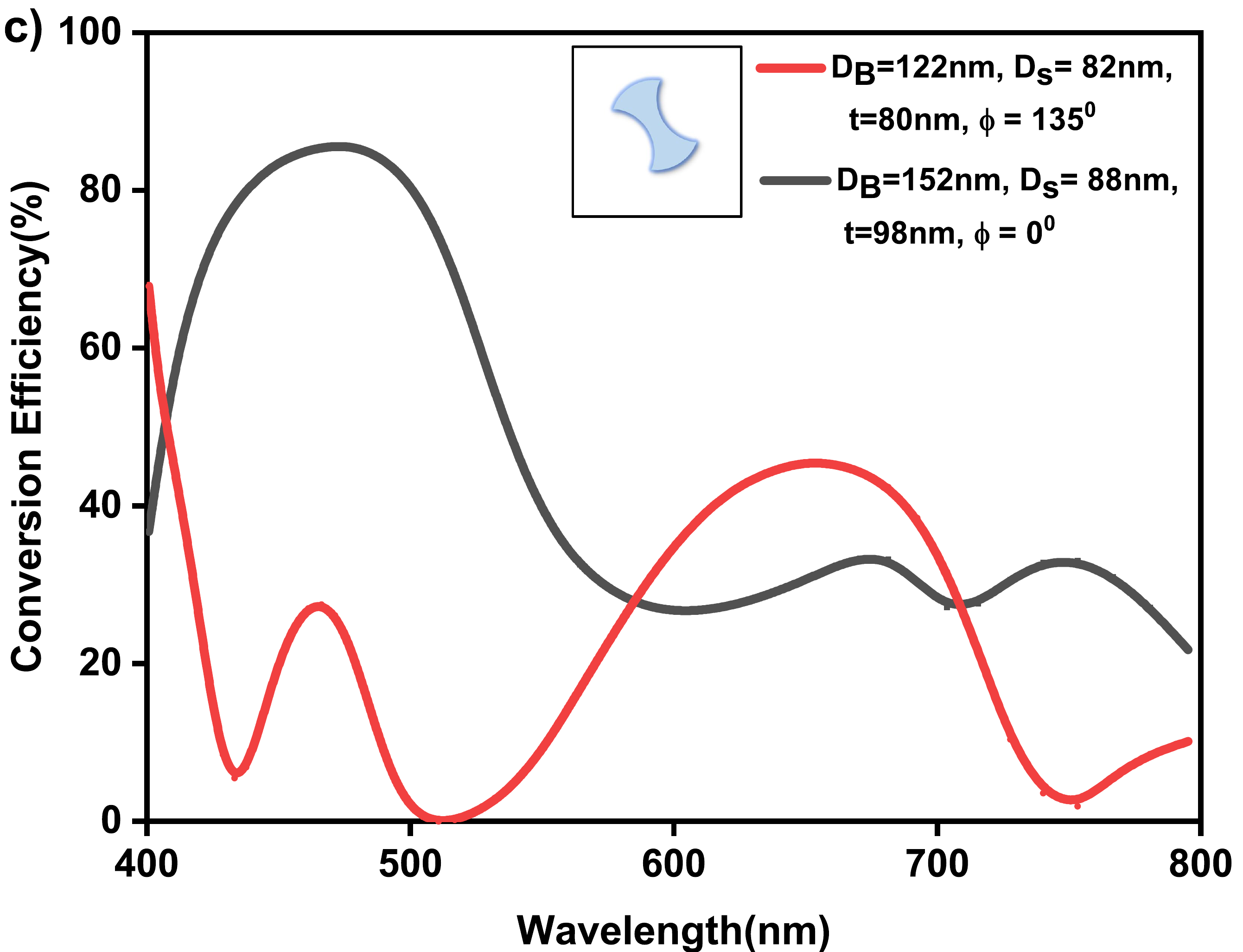} 
\includegraphics[width=6.7cm]{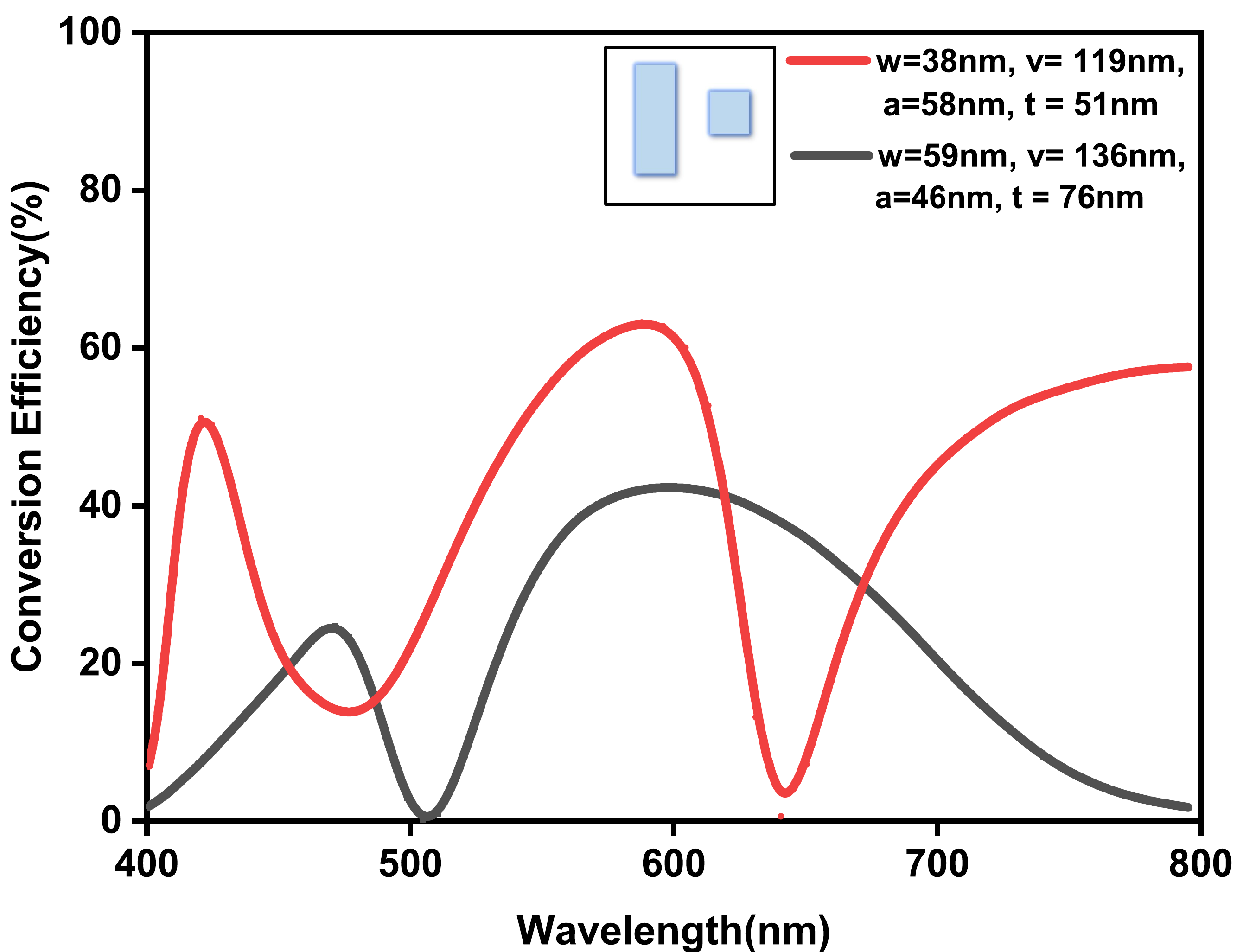} 
\caption{Gap-plasmon based half-wave plate metasurface: (a) A schematic illustration of unit cell of gap-plasmon based half-wave plate. (b) Different Al-nanoantenna geometrical designs -- meta-atoms (rectangle, double-arc) and meta-molecules (rectangle-circle pair, rectangle-square pair) with variable structural parameters. (c) Simulated conversion efficiency spectrum for double arc meta-atom and rectangle-square pair meta-molecule. Red and black curves shows different values for parameters in same design class.}
\end{figure}

\newpage
\begin{figure}[h!]
\centering \includegraphics[width=14cm]{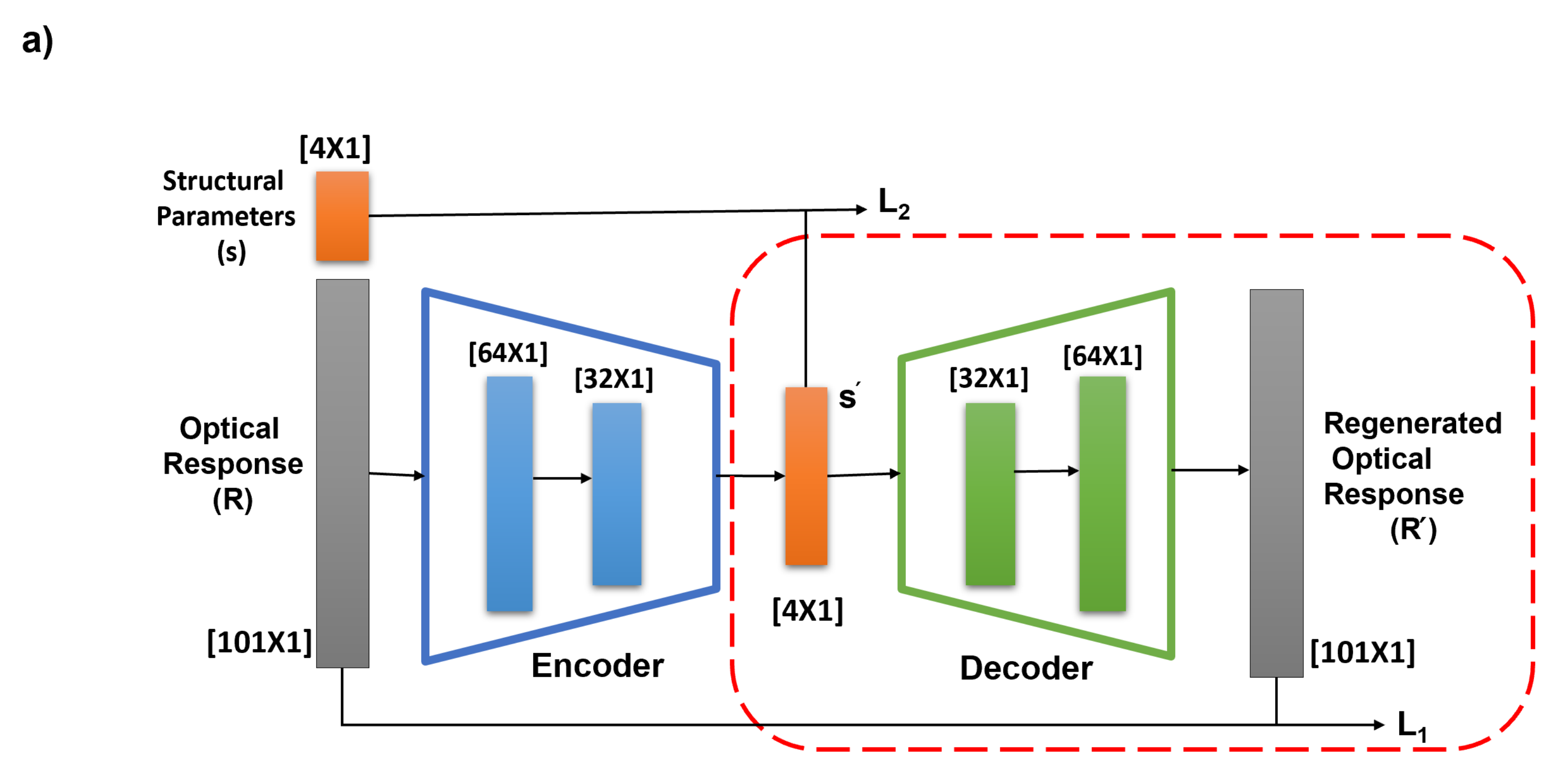}
\centering \includegraphics[width=14cm]{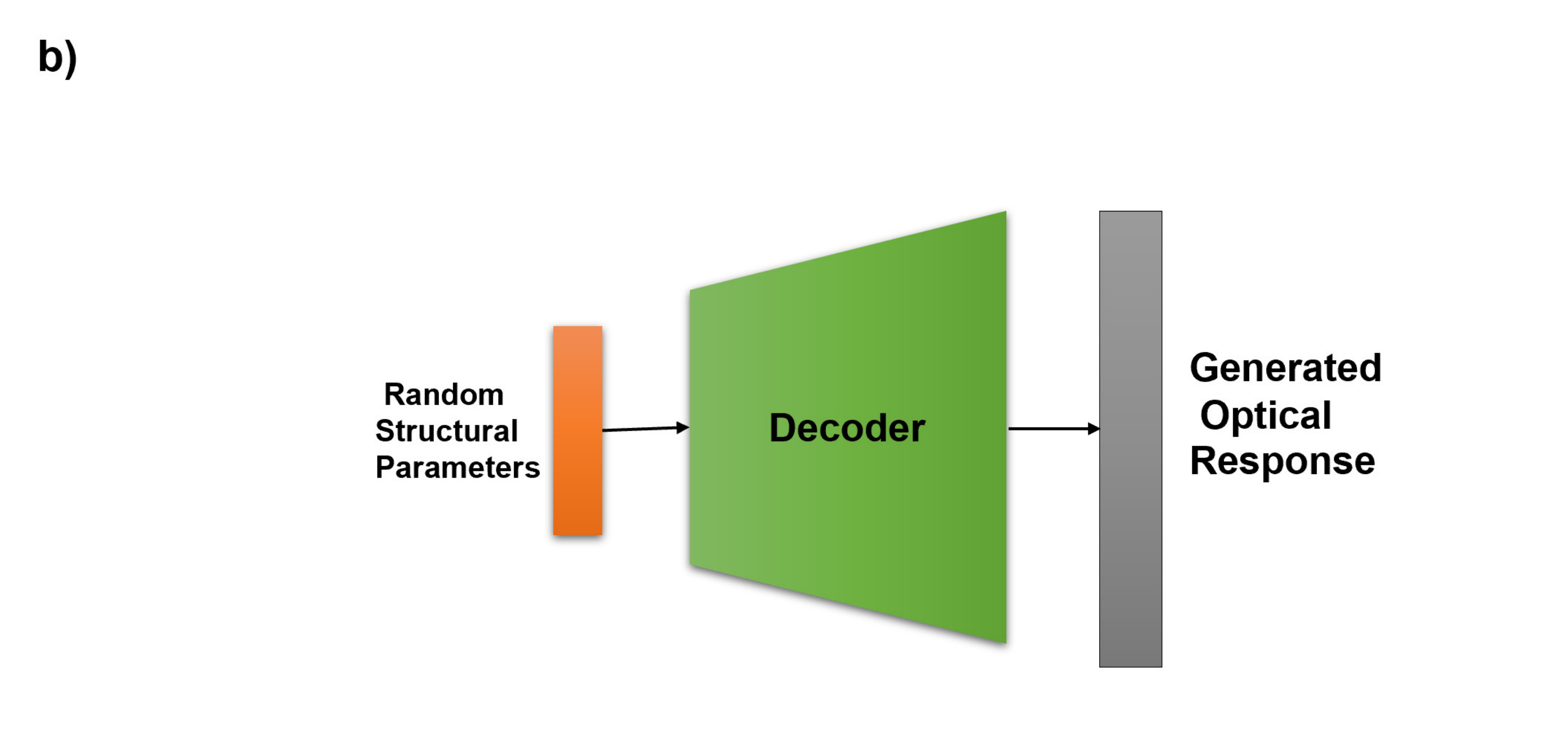}
\caption{Schematic of the biAE architecture: (a) In the training process, the encoder accepts the optical response and produces structural parameters as latent vectors. The decoder then regenerates the optical response from structural parameters. (b) After the training, the decoder (red box) in (a) can be separately used as a generator of optical responses for randomly selected structural parameters set $s = [L_x, L_y, \phi, t], [D_s, D_B, \phi, t], [w, v, d, t] \text{ or } [w, v, a, t]$ }
\end{figure}

\newpage
\begin{figure}[h!]
\includegraphics[width=6.75cm]{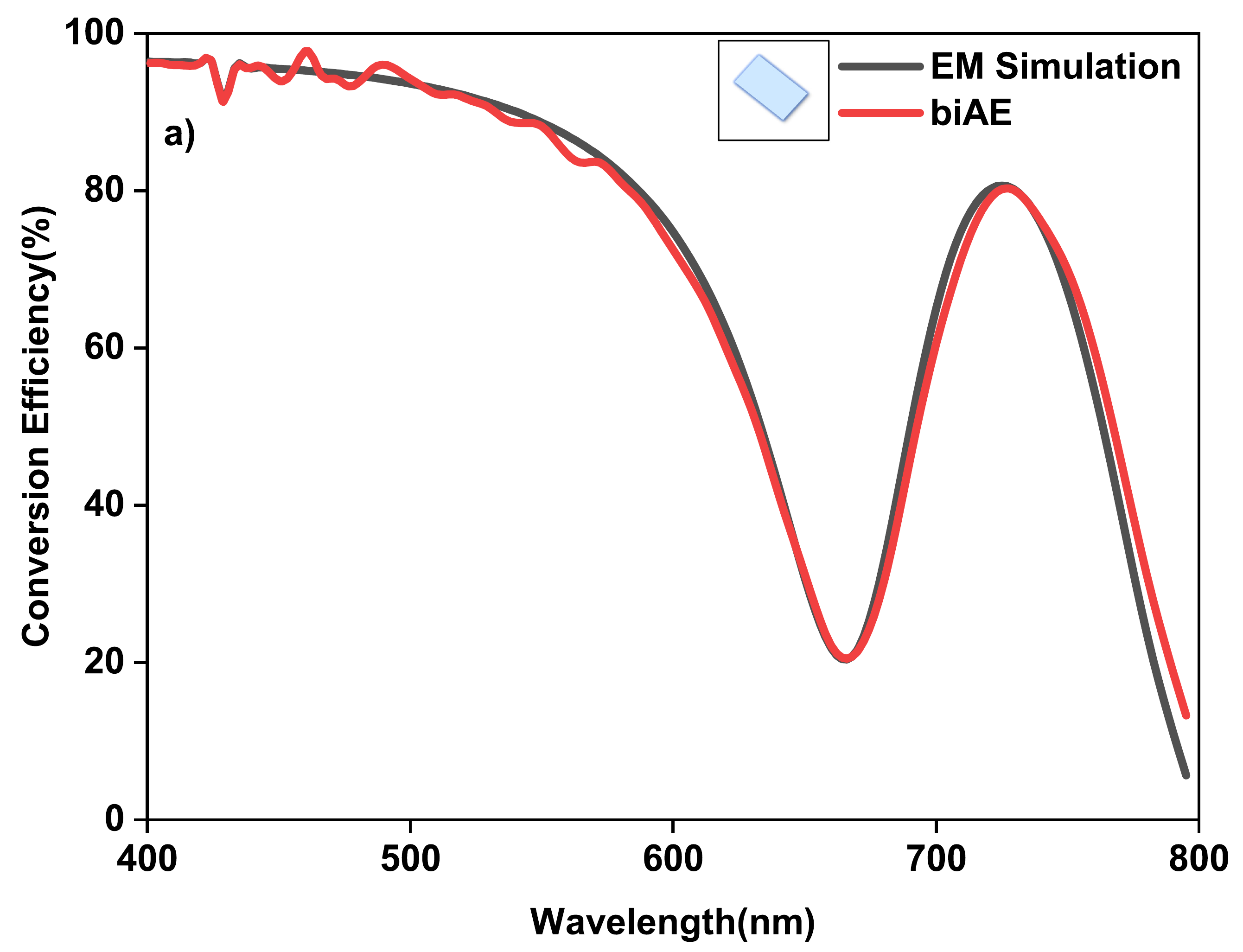} 
\includegraphics[width=6.75cm]{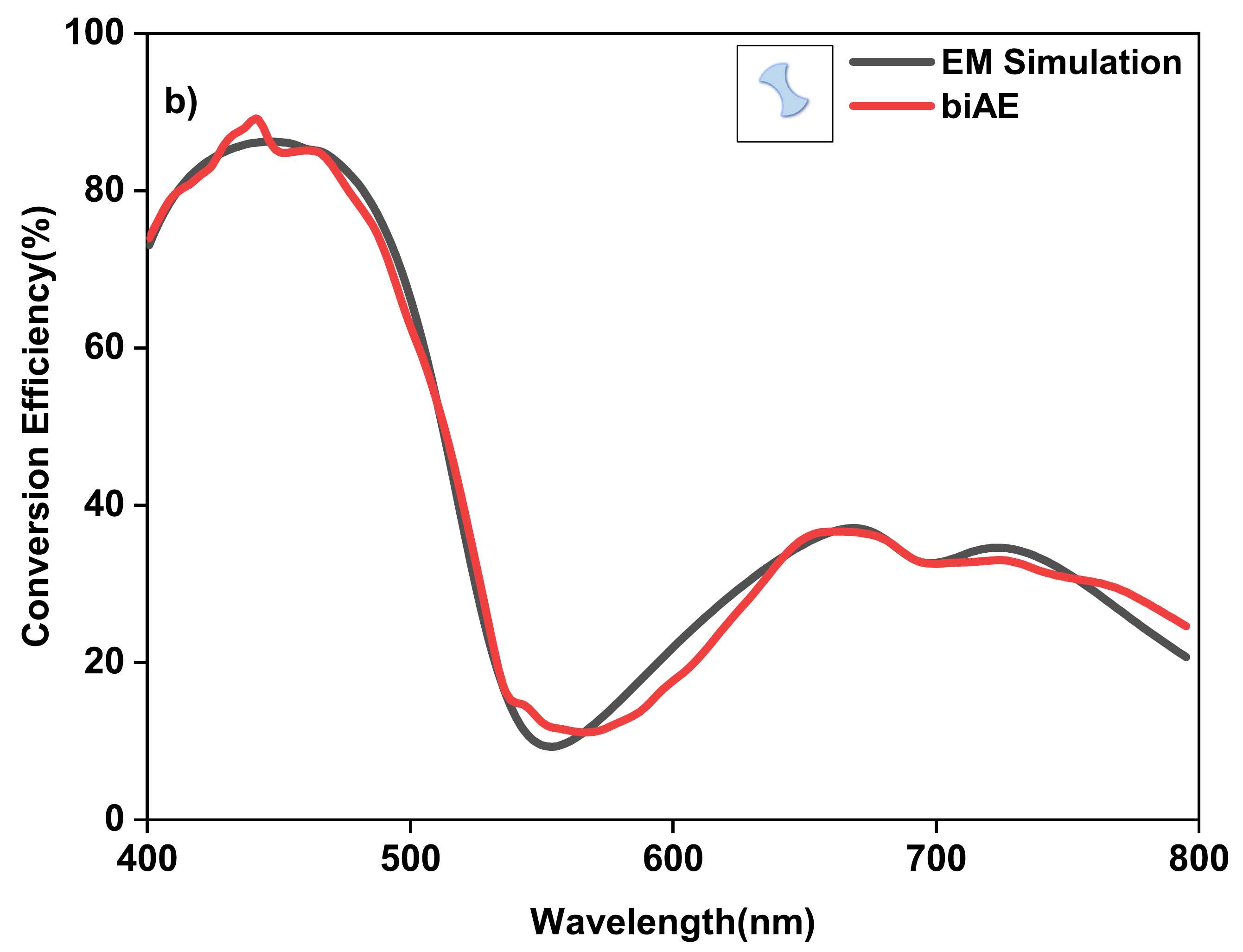} 
\includegraphics[width=6.75cm]{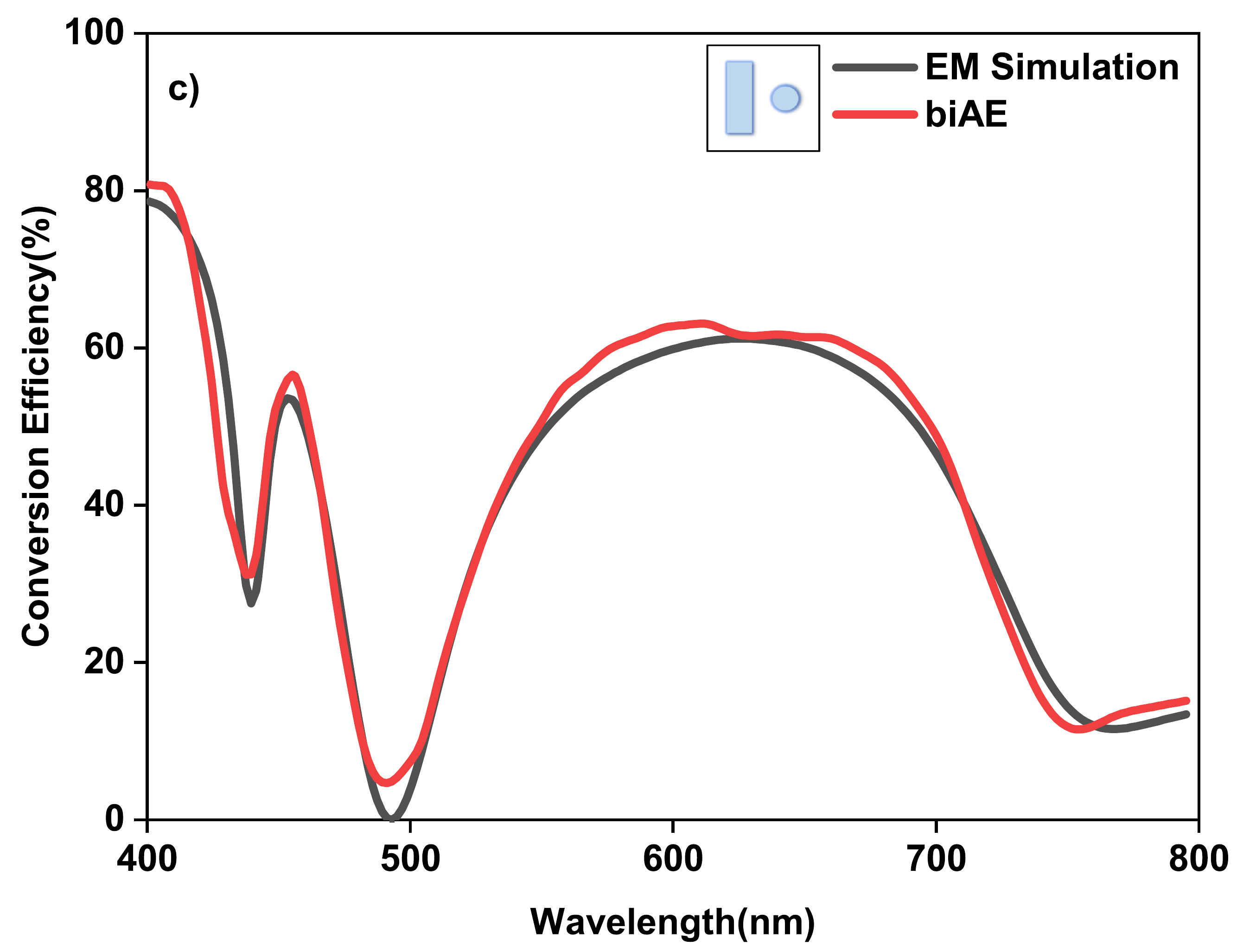} 
\includegraphics[width=6.75cm]{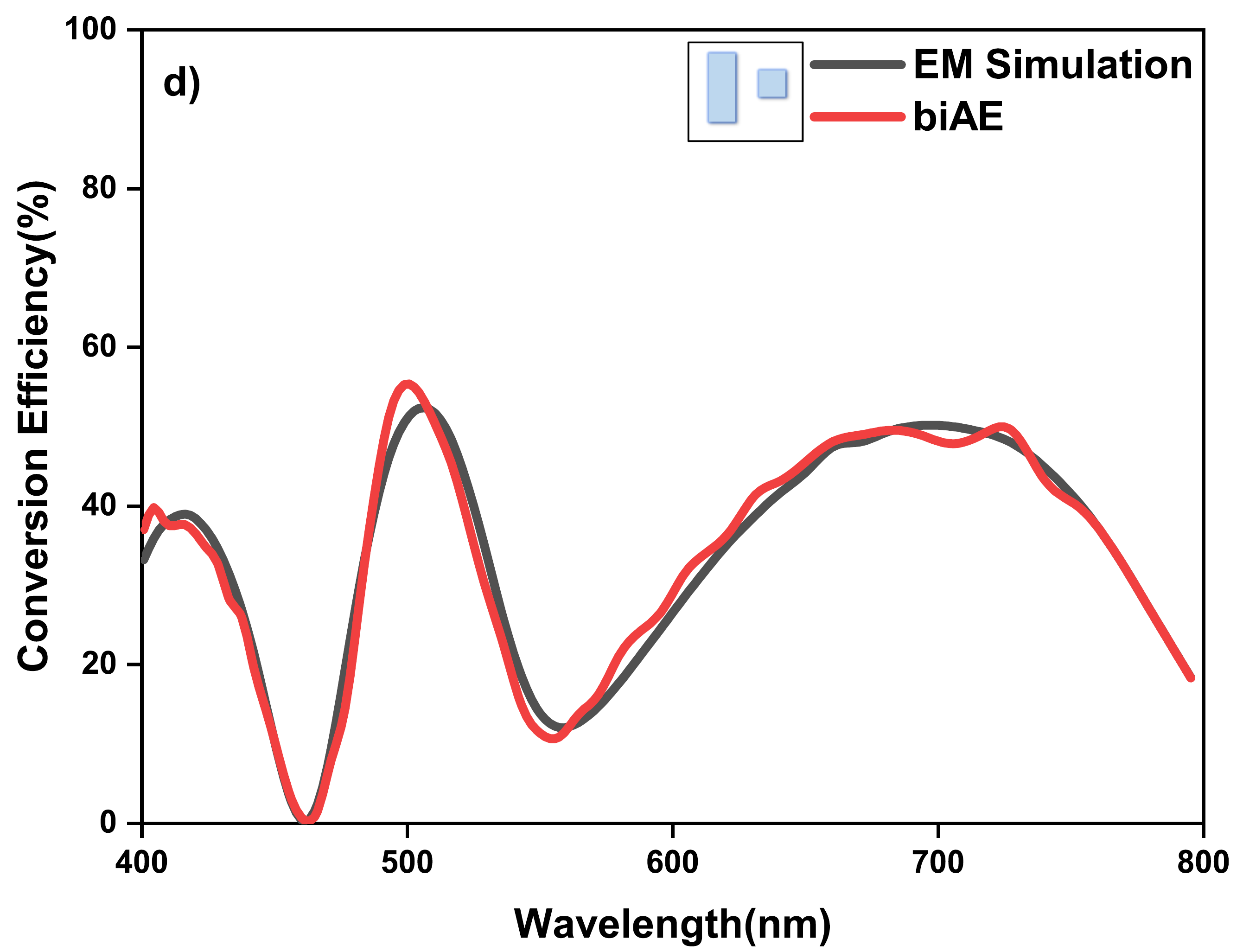} 
\caption{Comparison of biAE regenerated (red curve) and EM simulated (black curve) conversion efficiency for different class of geometrical designs: (a) Rectangle, (b) Double arc, (c) Rectangle-circle pair, and (d) Rectangle-square pair, demonstrating excellent biAE accuracy.}
\end{figure}

\newpage

\begin{figure}[h!]
\includegraphics[width=6.75cm]{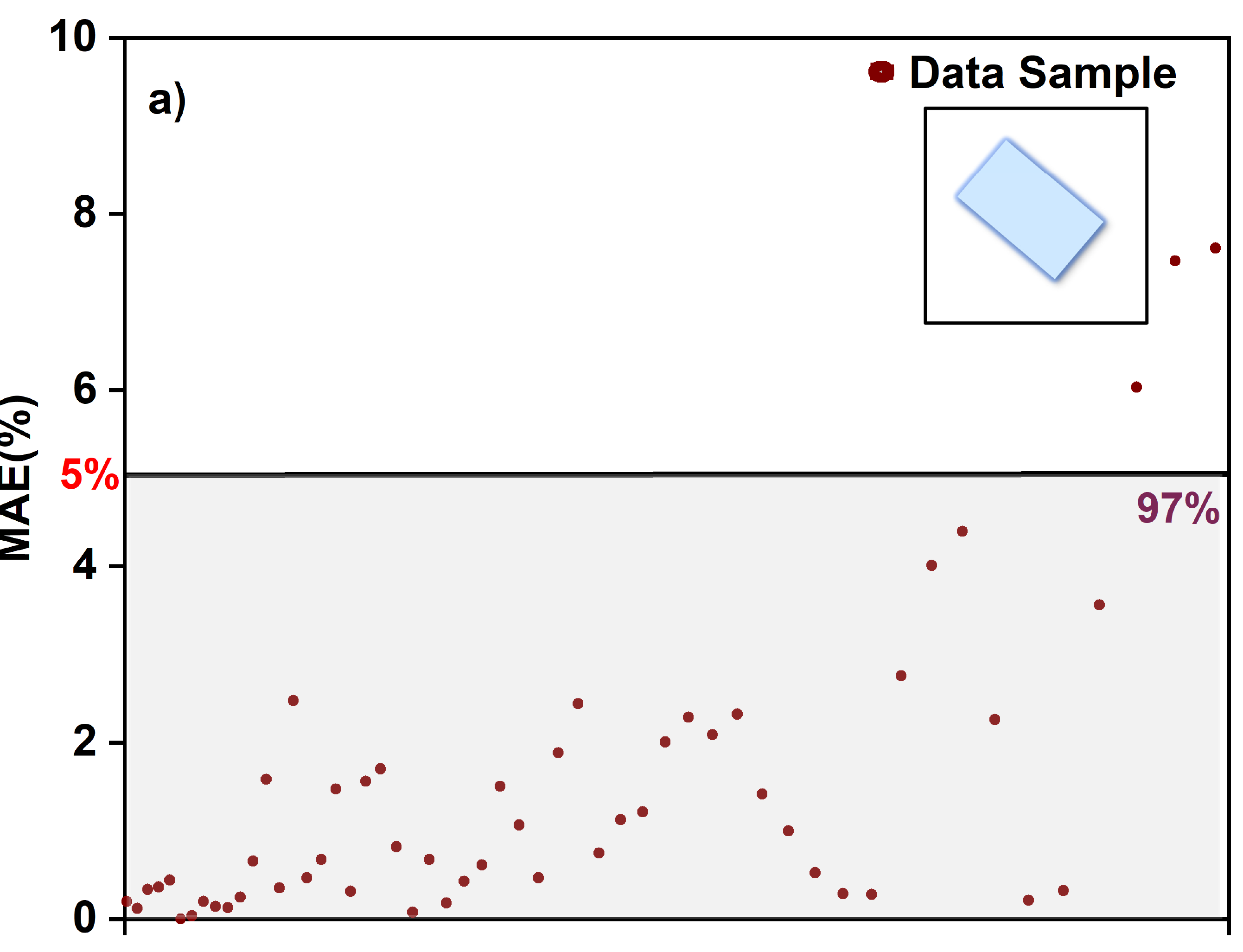} 
\includegraphics[width=6.75cm]{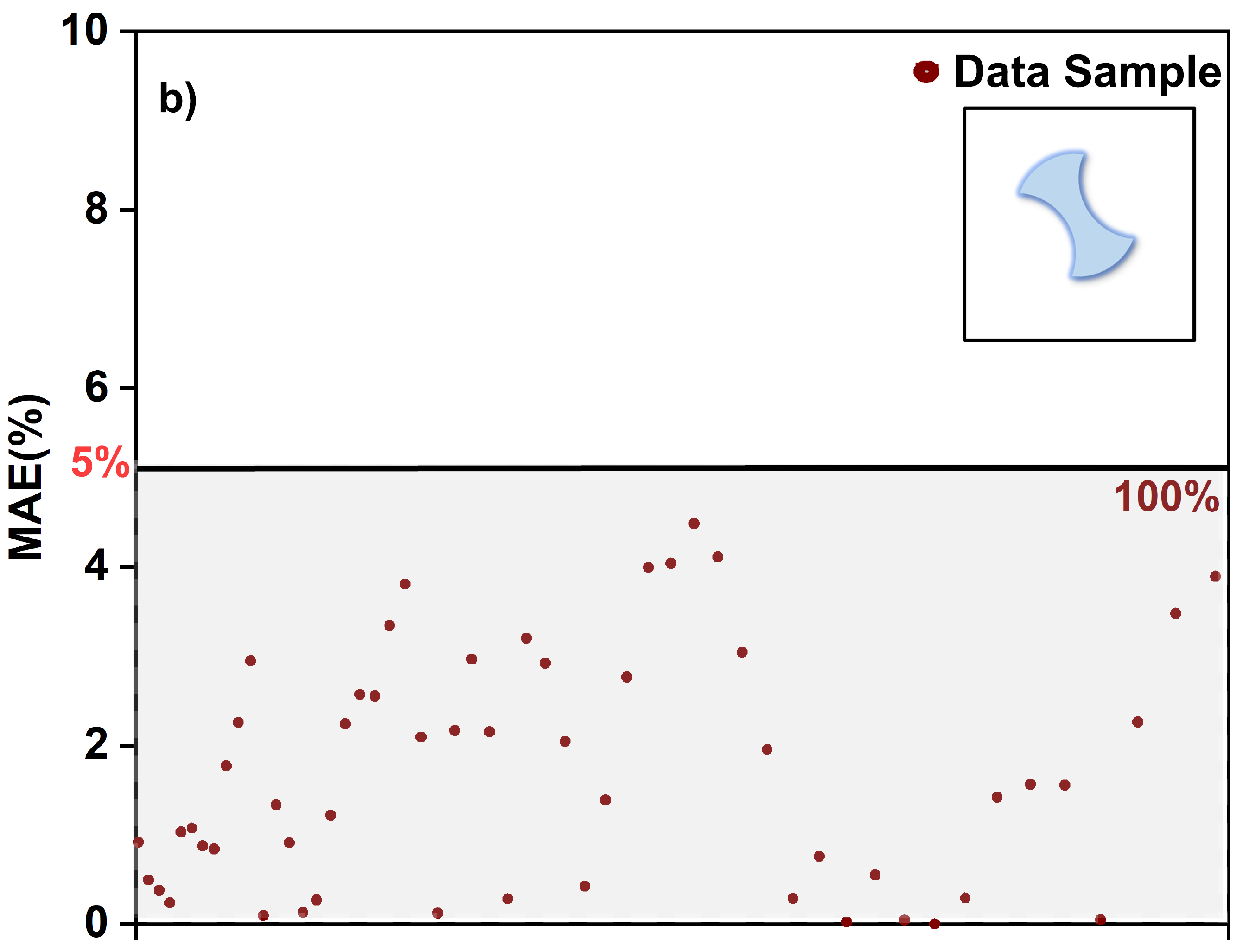} 
\includegraphics[width=6.75cm]{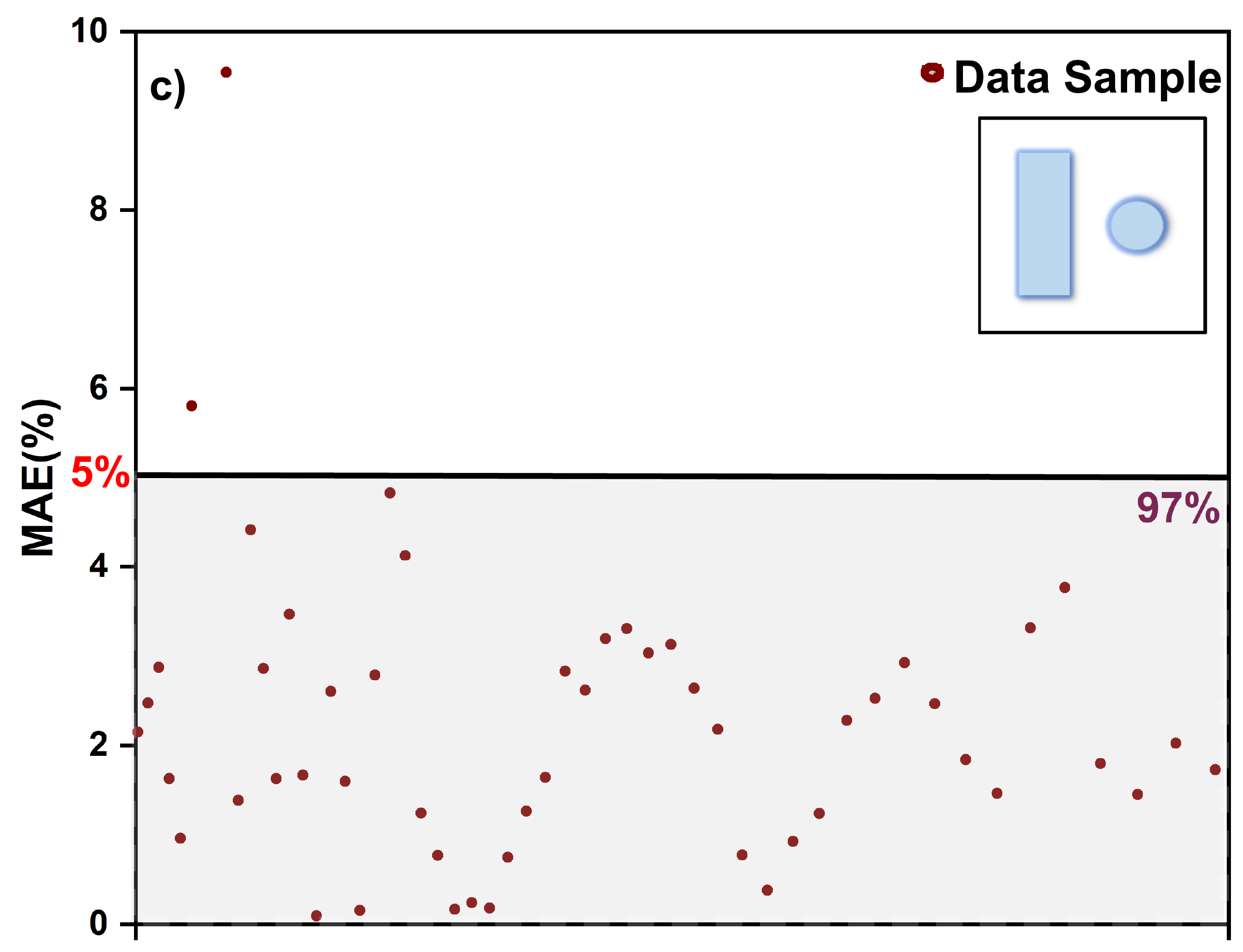} 
\includegraphics[width=6.75cm]{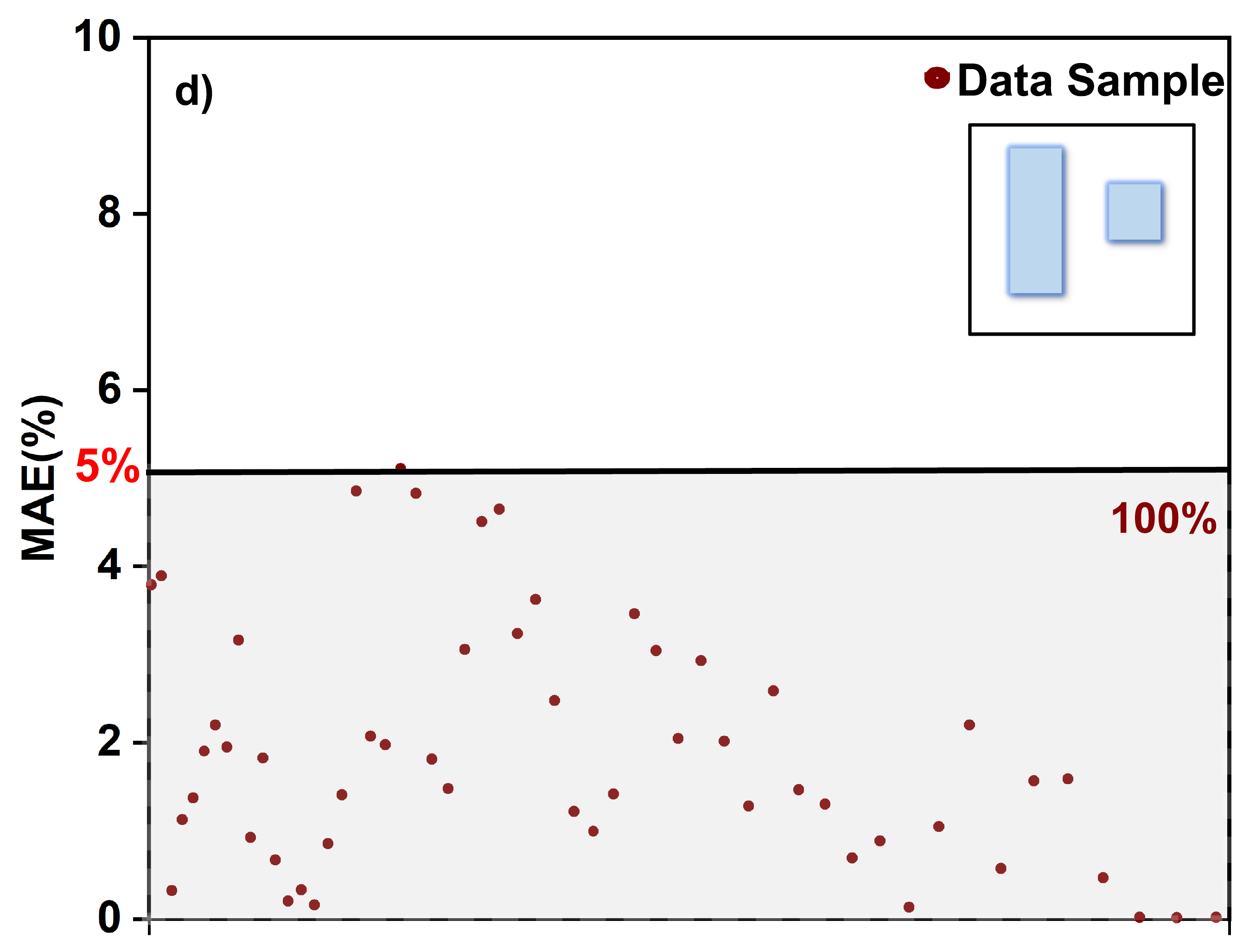} 
\caption{Average MAE on validation dataset for varying structural parameters for (a) rectangle, (b) double arc, (c) rectangle-circle pair, and (d) rectangle-square pair geometrical designs. }
\end{figure}

\newpage
\begin{figure}[h!]
\includegraphics[width=15cm]{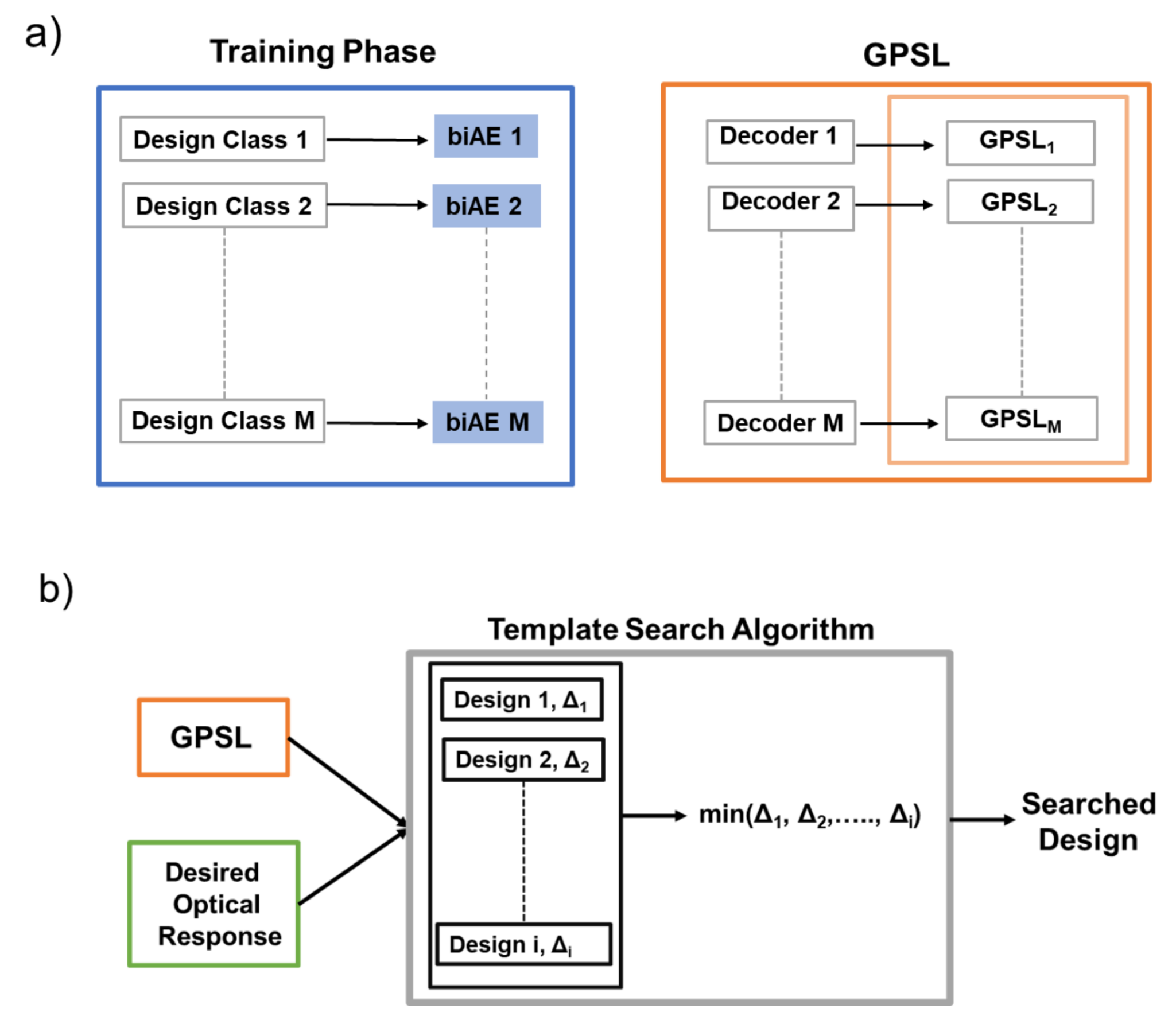} 
\caption{Schematic of generation of GPSL and working of template search algorithm. (a)  For each design class a biAE is trained. All the biAE have same model architecture and loss. The trained decoder network of biAE is used to generate GPSL. GPSL consists of sub-libraries (GPSL$_1$, GPSL$_2$, ...., GPSL$_M$) corresponding to each geometrical design class. Here we have four design classes, hence M=4 in our case (b) the template search algorithm outputs design with least MAE from a list of geometrical designs and having similar optical response for a given desired optical response as input.}

\end{figure}

\newpage
\begin{figure}[h!]
\includegraphics[width=17cm]{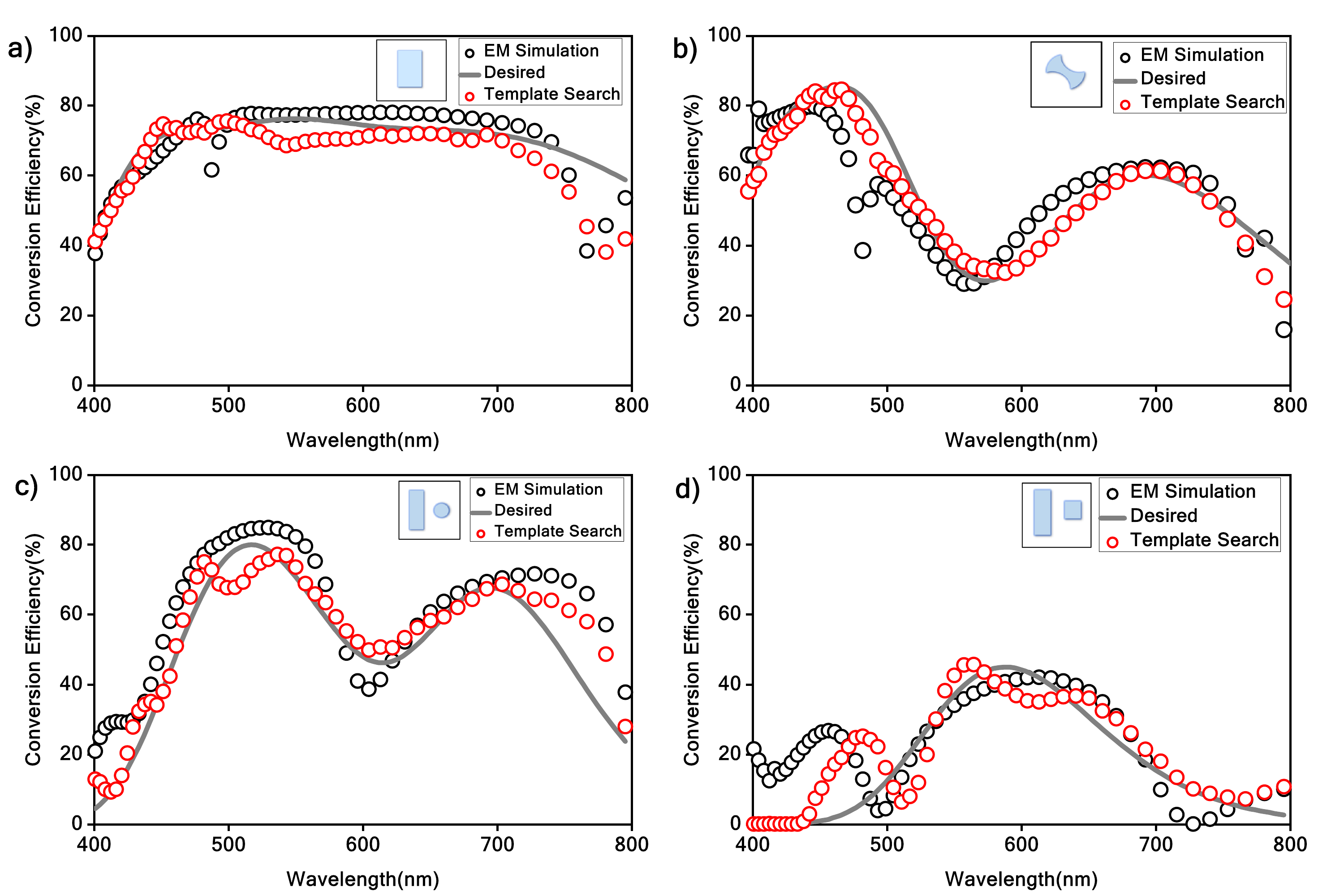} 
\caption{On-demand design of HMs for a desired optical response: The template search algorithm predicts (a) rectangle, (b) double arc, (c) rectangle-circle pair and (d) rectangle-square pair for different mixture of Gaussian as desired optical response as a single-band, dual-band, and broadband HMs device.}
\end{figure}
 
\newpage

\begin{figure}[h!]
\centering \includegraphics[width=11cm]{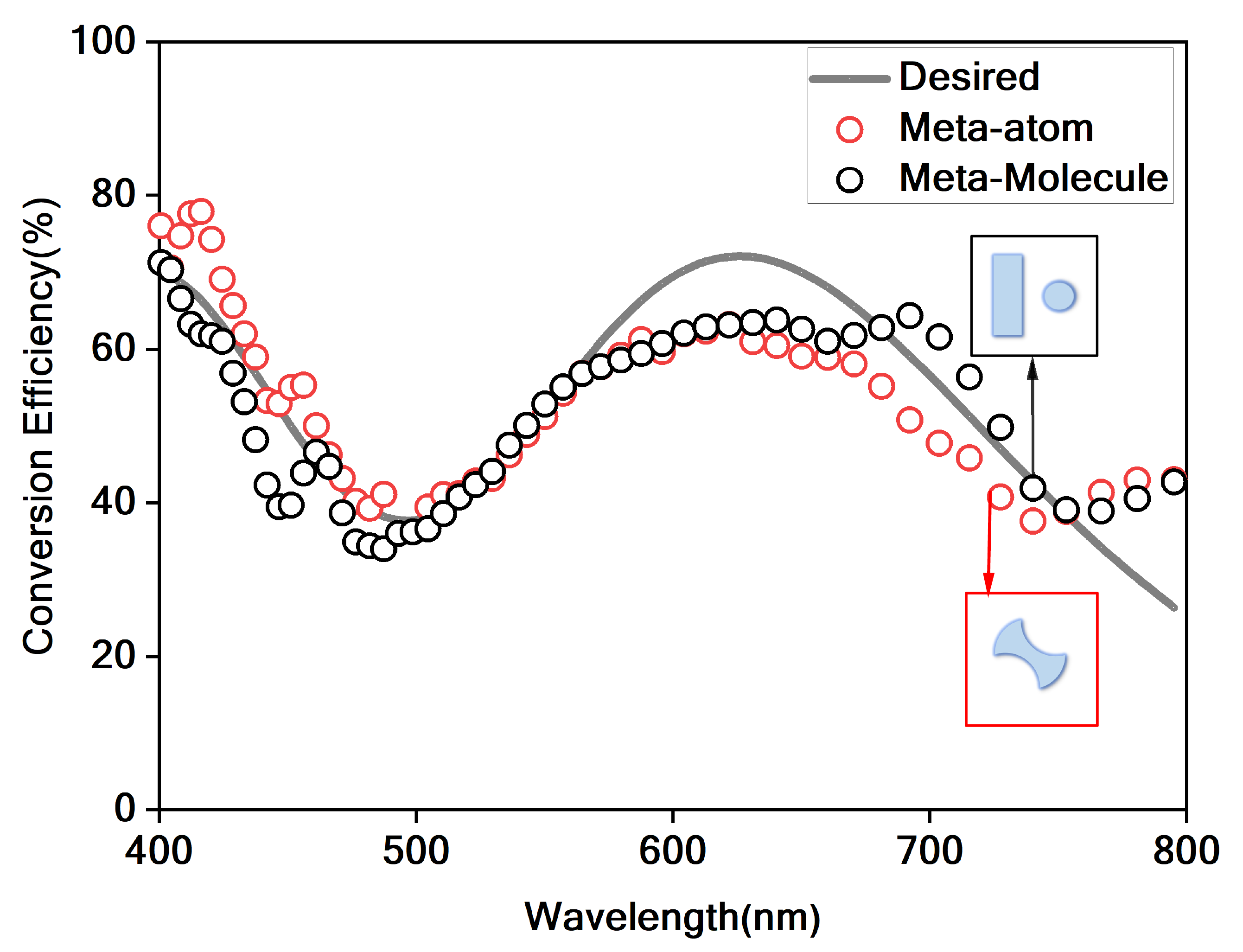} 
\caption{One to many mapping for a desired optical response: Template search methodology searches double arc as meta-atom and rectangle-circle pair as meta-molecule for same conversion efficiency optical response with MAE $\sim$2.5\%}.
\end{figure}

\newpage
\bibliography{references.bib}

\end{document}